# Phase-shifting coronagraph

François Hénault, Alexis Carlotti, Christophe Vérinaud
Institut de Planétologie et d'Astrophysique de Grenoble
Université Grenoble-Alpes, Centre National de la Recherche Scientifique
B.P. 53, 38041 Grenoble – France

## ABSTRACT

With the recent commissioning of ground instruments such as SPHERE or GPI and future space observatories like WFIRST-AFTA, coronagraphy should probably become the most efficient tool for identifying and characterizing extra-solar planets in the forthcoming years. Coronagraphic instruments such as Phase mask coronagraphs (PMC) are usually based on a phase mask or plate located at the telescope focal plane, spreading the starlight outside the diameter of a Lyot stop that blocks it. In this communication is investigated the capability of a PMC to act as a phase-shifting wavefront sensor for better control of the achieved star extinction ratio in presence of the coronagraphic mask. We discuss the two main implementations of the phase-shifting process, either introducing phase-shifts in a pupil plane and sensing intensity variations in an image plane, or reciprocally. Conceptual optical designs are described in both cases. Numerical simulations allow for better understanding of the performance and limitations of both options, and optimizing their fundamental parameters. In particular, they demonstrate that the phase-shifting process is a bit more efficient when implemented into an image plane, and is compatible with the most popular phase masks currently employed, i.e. four-quadrants and vortex phase masks.

## 1  INTRODUCTION

With the recent commissioning of ground instruments such as SPHERE [1] or GPI [2] and future space observatories like WFIRST-AFTA [3], coronagraphy should probably become the most efficient tool for identifying and characterizing extra-solar planets in the forthcoming years. But this type of instruments are subject to extremely severe image quality requirements: for example, achieving a contrast ratio of $10^{-6}$ suitable to the detection of hot giant planets typically requires a Wavefront error (WFE) better than 10 nm RMS. Hence they must be assisted with extreme adaptive optics (XAO) systems, either part of the host telescope or of the instrument itself. In addition, there is a strong interest to limit the so-called Non common path aberrations (NCPA), i.e. differential WFE arising between the science and wavefront sensing optical trains, from the time they are separated (usually by means of a beam splitting plate). Hence coronagraphs shall include internal Wavefront sensors (WFS) located as close as possible to the science image plane in order to compensate for those NCPA. Practically however, the separation often occurs at the entrance of the coronagraph, for example by means of a small mirror reflecting starlight to the WFS while planet light enters the coronagraph. Moreover, the employed WFS often are of the low-order type (LOWFS) and limited to a few Zernike modes.

In this communication is investigated the ability of phase-shifting techniques to overcome the previous limitations, by enabling more accurate WFE measurements and pushing back the separating element as far as possible behind the coronagraphic mask. Before applying this concept to coronagraph observations, we firstly review the existing designs of phase-shifting telescopes in section 2. Their mathematic formalism adapted to phase-shifting coronagraphs is described in section 3. In section 4 are presented the main results of extensive numerical simulations where phase-shifts may be introduced in both pupil and image planes, and the critical system parameters have been optimized. A conceptual optical implementation of both concepts is proposed in section 5. Finally, the main conclusions are drawn in section 6.

## 2 A BRIEF HISTORY OF PHASE-SHIFTING TELESCOPES

Before developing the theory of phase-shifting coronagraphs in section 3, it is worth reviewing the concept of Phase-shifting telescope (PST), where the phase-shift process was originally proposed to sense the optical aberrations or atmospheric disturbance degrading the performance of astronomical telescopes. Four of these concepts can arbitrarily be distinguished and are illustrated in Figure 1.

### 2.1 Mach-Zehnder telescope interferometer (MZTI)

It seems that the first proposal of a phase-shifting process for sensing the static or dynamic phase errors of an astronomical telescope originates from R. Angel and J. L. Codona [4-5]. The basic concept is illustrated in Figure 1-1, where it appears as a Mach-Zehnder interferometer installed at the focal plane of the telescope. The incoming beam is firstly separated into a main arm and a reference arm by a beam splitting plate as indicated in the figure. The reference beam is then filtered spatially with a pinhole, cleaning the WFE and generating a spherical reference wavefront. In the mean time, phase shifts are introduced along the main arm, for example by axial displacement of one mirror. Those phase-shifts are typically equal to $\phi = 0$, $\pi/2$, $\pi$ and $3\pi/2$, though many other sequences are possible. After beam recombination on the second beamsplitter plate, the beams are imaged onto a pupil plane where an interferometric fringe pattern is formed. The exit telescope wavefront can then be reconstructed using classical algorithms of phase-shifting interferometry. Recently, this concept was chosen as the baseline design for wavefront sensing inside the GPI instrument [6]. It must be noticed however that:

- The MZTI is particularly well suited to Lyot-type coronagraphs where the starlight is blocked by a central mask located in the image plane (e.g. the apodized pupil Lyot coronagraph of the GPI instrument [6]). Apparently, no application to other types of coronagraphs has been reported.

- To our knowledge, its theory has always been presented under the hypothesis of weak aberrations, where an aberrated wave $\exp[i\,kW]$ is approximated as $1 + i\,kW$, with $W$ the wavefront, $k$ the wavenumber and $i$ the square root of $-1$. Referring to the basic principle of phase-shifting interferometry, it seems however that the assumption is not necessary.

- Finally, it should be noted that its optical design is somewhat complex, bulky and difficult to align, all drawbacks which do not affect the following concepts.

### 2.2 Off-axis telescope interferometer (OATI)

Proposed in 2005 [7], the OATI concept may be considered as a good alternative to the MZTI, since reference phase-shifts are not introduced sequentially with the help of a driving mechanism, but by decentering the reference arm with respect to the main optical axis, thus producing a stationary fringe pattern in the image plane. The optical concept is depicted in Figure 1-2 and basically consists in adding auxiliary optics located in front or besides the main telescope, then focusing starlight onto a Single mode optical fiber (SMF). On its exit side, the SMF generates a reference spherical wavefront that is mixed with the main wavefront in the telescope image plane, where it arrives under an angle α ad generates a weak fringe pattern. If the angle α and the ratio $\rho$ between the exit pupil diameters of the main and reference beams fulfil certain conditions, the input wavefront can be reconstructed from the fringe pattern using a Fourier transform algorithm [7]. Hence the sequential temporal phase-shifts of the MZTI are replaced with a static spatial phase-shift generated by the tilted reference wavefront. The OATI concept has later been adapted to the case of a Four-quadrant phase mask (FPQM) coronagraph, where the reference beam is introduced behind the phase mask (see for example Ref. [8] and the cited reference therein).

However the OATI concept presents a fundamental limit that was evidenced in later publications [9-10]. It results from the intrinsic chromaticity of the fringe patterns generated by the off-axis reference arm, whose inter-fringe distance is proportional to λ/α, with λ the central wavelength of the considered spectral band. Studies reported in Refs. [9-10] demonstrate that the resulting losses in Signal-to noise ratio (SNR) are so critical that this concept should be abandoned in favor of a Phase-shifting telescope interferometer (PSTI) described in the next sub-section. Finally, it should be that technical solutions exist to removing the intrinsic chromaticity from the apparatus, but only at the price of significant complication in the optical design (see for example the description of the FANI concept in Ref. [11]).

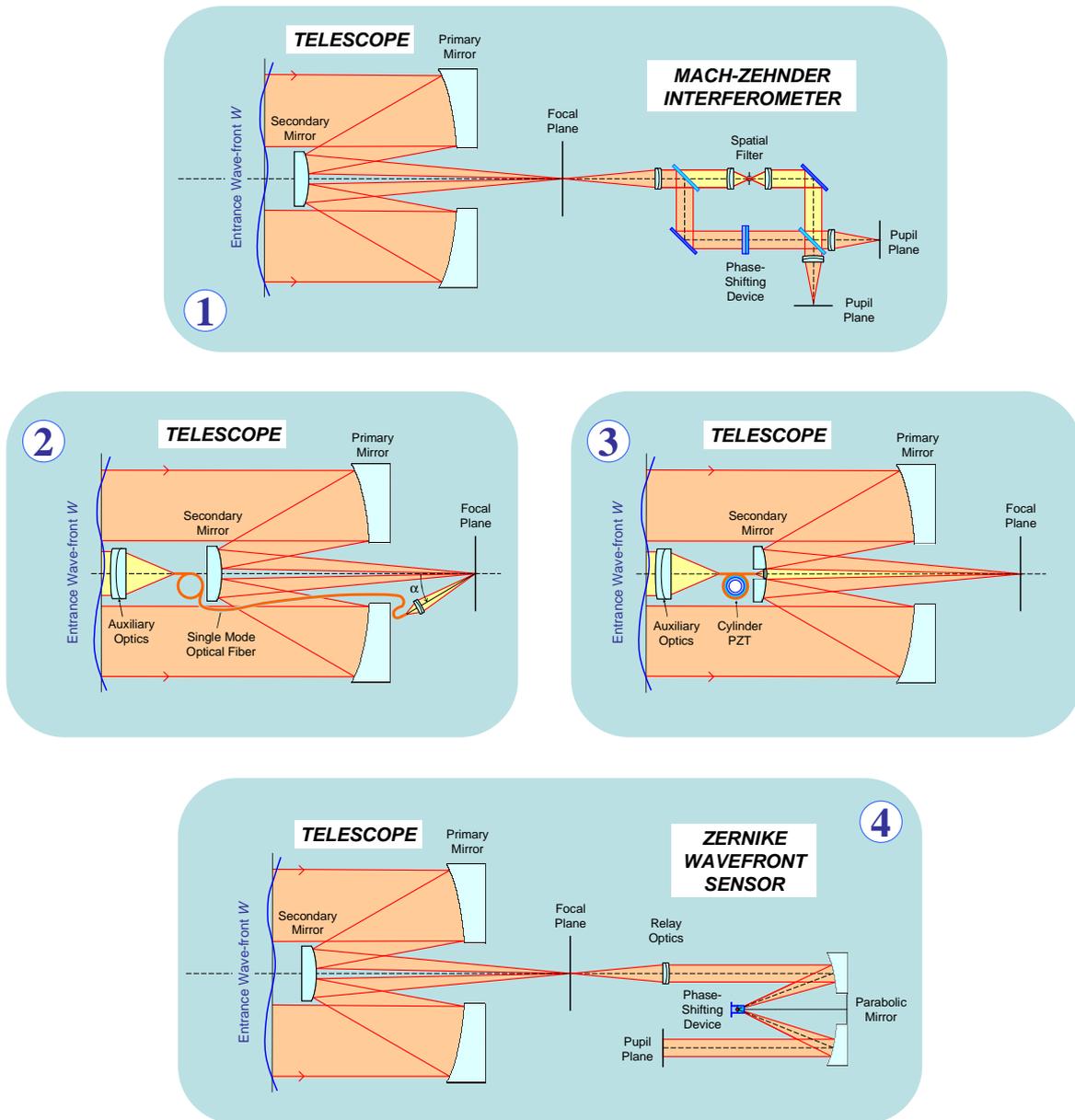

Figure 1: Four different concepts of phase-shifting telescopes. It should be noted that concepts n° 1 and 4 are pupil plane wavefront sensors, while concepts n° 2 and 3 are image plane wavefront sensors.

### 2.3 Phase-shifting telescope interferometer (PSTI)

This concept was proposed in 2006 [12] in order to combine the advantages of both the MZTI and OATI (simple optical scheme, no fringe chromatism). It is illustrated in Figure 1-3 in one of its simplest arrangement. It again comprises auxiliary optics injecting starlight into a SMF, but here the fiber exit is placed on-axis, for example passing through the central obscuration on the secondary mirror of the telescope. Successive phase-shifts $\phi$ are added into the reference arm by means of a cylindrical Piezo-electric transducer (PZT). The number of phase-shifts can be reduced to three, with $\phi = 0$, $2\pi/3$ and $4\pi/3$.

Like the previous OATI concept, the PSTI is an image plane wavefront sensor. The three PSF recorded in the focal plane are processed mathematically to reconstruct the input wavefront from the telescope, as summarized in § 3.2. The optical concept is simple and totally immune to inter- fringe chromatism, as the reference arm is centered on the main optical axis. Furthermore, it is not limited to weak aberrations. It has later been proposed for co-phasing the segmented mirrors of future giant telescopes [13-14], or in the field of stellar interferometry for imaging [15] or fringe tracking purposes [16]. Numerical simulations demonstrated that the concept is efficient for all those cases. Moreover, it has recently been applied to a vortex coronagraph and tested at the Palomar Observatory [17].

### 2.4 Phase-shifting Zernike wavefront sensor (PS-ZWFS)

The most recent phase-shifting concept is perhaps also the most promising: It has been proposed by Wallace *et al* in 2011 [18] and is schematically illustrated in Figure 1-4. Its great originality is to introduce phase-shifts into a small circular area of the image plane (typically smaller than the Airy disk) and to sense intensity variations in the pupil plane. This can be achieved for example by means of an optical fiber having a reflective tip and axially driven by a PZT, as described in the original publication. It must be noticed that this concept will be integrated into the next adaptive optics system equipping the 5-m Mount Palomar telescope [19].

As the very first design of MZTI, the PS-ZWFS actually is a pupil plane WFS, and its analysis was only presented in the theoretical frame of weak aberrations. Moreover; it seems that its application to coronagraph instruments has not been studied yet.

### 2.5 Conclusion

At the term of this quick tour of PST concepts, it seems that the most eligible for coronagraphic applications are the two last ones, i.e. the phase-shifting telescope interferometer and the phase-shifting Zernike WFS described in § 2.3 and 2.4 respectively. However a few questions remain open at this stage:

- o Do the concepts still work when integrated into a coronagraph, and preferably after its diffractive component ? What are their critical parameters ?
- o Are they limited to weak aberration measurements ? Are they solely low-order WFS or could they sense higher order spatial frequencies ?

The goal of the following sections is to provide the reader with some elements of answer, firstly by theoretical analysis (§ 3), then with the help of numerical simulations (§ 4).

## 3 THEORY OF PHASE-SHIFTING CORONAGRAPH

After reviewing the basic principle of a coronagraph and defining the employed mathematical notations and coordinate frames (sub-section 3.1), the theoretical relationships applicable to three different phase-shifting processes are established: phase-shifts can either be introduced into the telescope pupil plane (sub-section 3.2), in the Lyot stop plane of the coronagraph (sub-section 3.3), or in the plane of the image mask or one of its optical conjugates (sub-section 3.4). Most analytic notations employed into the text and figures of the whole section are summarized in the nomenclature of Appendix A.

### 3.1 Mathematical notations and coordinate frames

**A. Coronagraph reference frames**

The general principle of a stellar coronagraph is summarized in Figure 2-1. Located at the exit port of an astronomical telescope, it is usually composed of the following elements:

- The entrance plane of the coronagraph is located at the exit pupil of the telescope. The attached reference frame is OXYZ, where O is the pupil centre and OZ the main optical axis of the system. Any point P in the OXY plane is defined by Cartesian coordinates $(x,y)$. In that plane the WFE to be measured is noted $W(x,y)$ or $W(P)$ in shorthand notation.

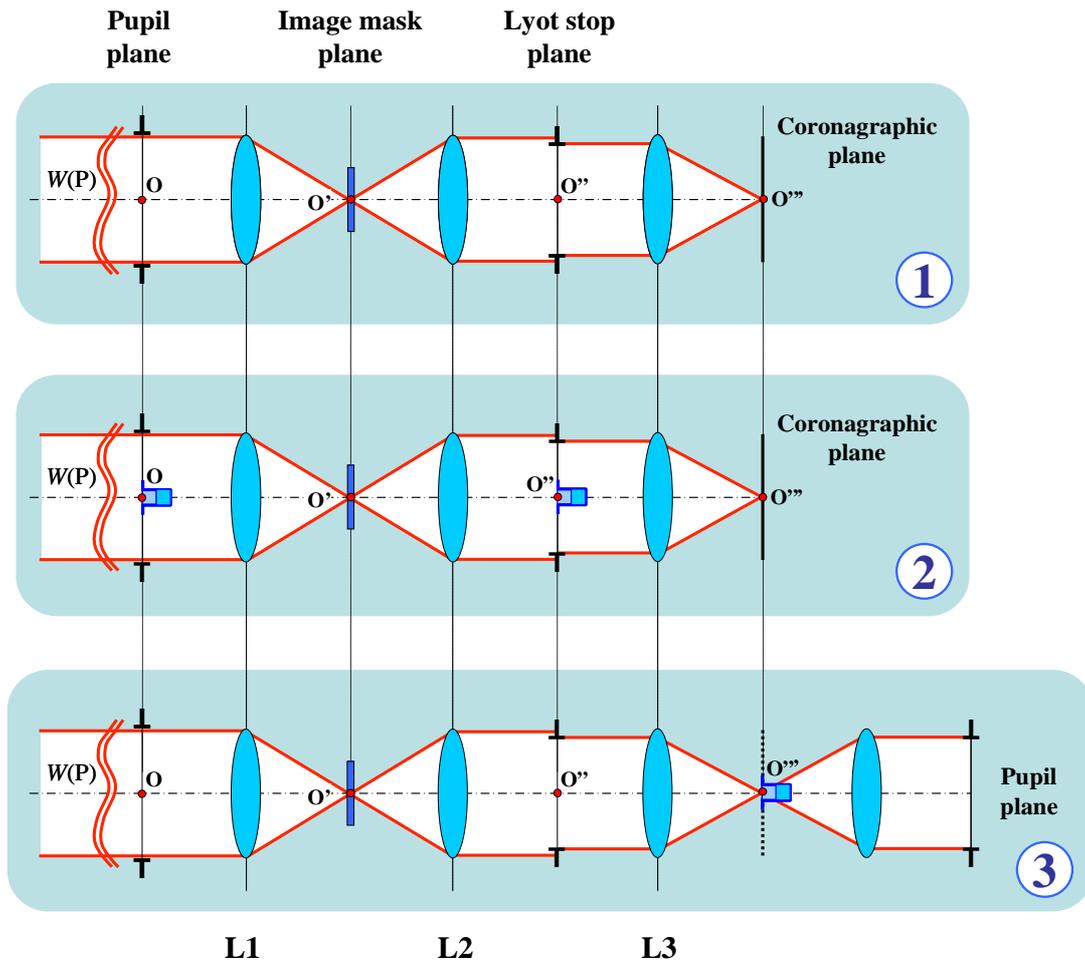
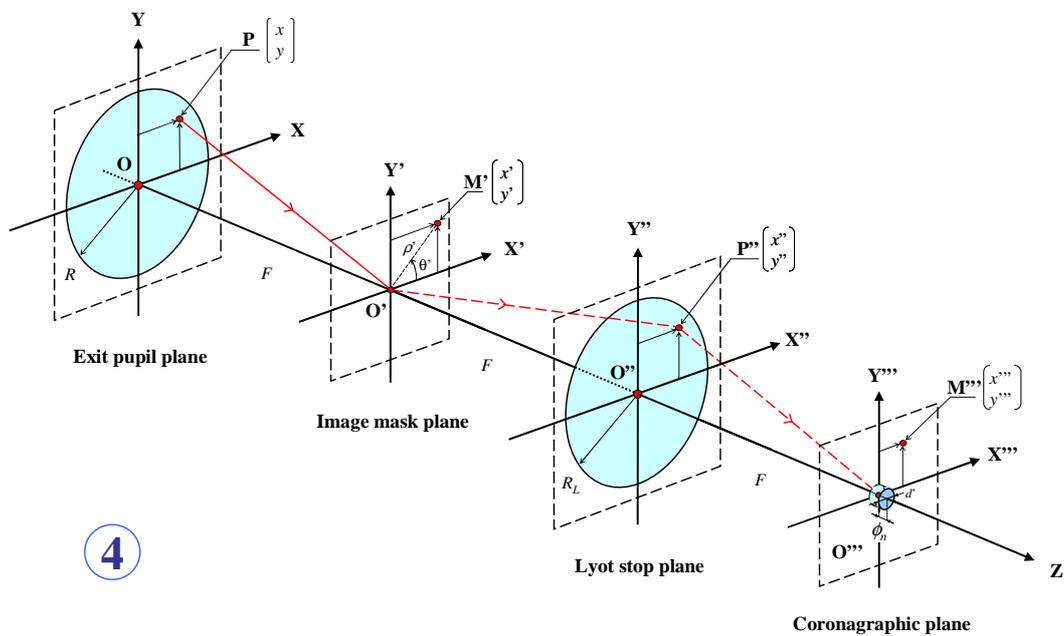

Figure 2: Mathematical notations and coordinate frames.

- The diffractive element is located in the image plane of the telescope. It can either be a transmission mask as in Bernard Lyot's historical design, or a phase mask in modern PMCs. The attached coordinate system is O'X'Y'Z, with O' the centre of the diffractive mask located along the main optical axis OZ. Any point M' in the O'X'Y' plane is defined by its Cartesian coordinates (x',y') or polar coordinates (ρ',θ'). The coronagraph is assumed to be a PMC with uniform transmission equal to unity. The phase mask function is noted ϕ(M') in shorthand notation.

- The Lyot stop is optically conjugated with the pupil plane OXY. The attached coordinate system is O"X"Y"Z, with O" the centre of the Lyot stop and (x",y") the Cartesian coordinates of any point P" in the O"X"Y" plane.

- The coronagraphic image plane is the final detection plane. The attached coordinate system is O'''X'''Y'''Z, with O''' the centre of the detector and (x''',y''') the Cartesian coordinates of any point M''' at the detector surface.

All employed coordinate systems are illustrated in Figure 2-4. The diameters of the telescope exit pupil and of the Lyot stop are noted $D = 2R$ and $D_L = 2R_L$ respectively. For the sake of simplicity, we assume identical focal length $F$ for the telescope and all focusing and collimating optics, and a magnification factor equal to unity between the input pupil and Lyot stop planes. All optics sketched as single lenses L1, L2 and L3 in Figure 2 could in reality be mirrors.

**B. Phase-shifts**

As presented in Figure 2-2 and 2-3 where they are symbolized with the  pictogram, phase-shifts shall be introduced at three different locations inside the coronagraph:

1) Into the exit pupil plane of the telescope OXY, over a small circular sub-pupil of diameter $d = \rho D = 2\rho R$ centered at point O, with $\rho$ the ratio between the reference and main pupil diameters. Intensity variations are recorded in the coronagraph image plane O'''X'''Y'''. This case is studied in § 3.2.

2) Into the Lyot stop plane O"X"Y", over the same sub-pupil centered at point O". Here also intensity variations are recorded in the coronagraph image plane. The case is studied in § 3.3.

3) Into the coronagraph image plane O'''X'''Y''', over a small circular area of diameter $d'$ centered at point O'''. Intensity variations are recorded into a pupil plane optically conjugated with the Lyot stop plane O"X"Y". The case is illustrated in Figure 2-3 and 2.4 and is studied in § 3.4.

As in Ref. [12] and in order to limit the number of required image acquisitions, we opt for a three-step phase-shift algorithm where successive phase-shifts $\phi_m$ are equal to $\phi_1 = 0$, $\phi_2 = 2\pi/3$ and $\phi_3 = 4\pi/3$. The measured intensity variations $I_m(P)$ at any point P of the detection plane can thus be expressed in the general form:

$$I_m(P) = A(P) + B(P)\cos\phi_m + C(P)\sin\phi_m \qquad \text{with:} \quad 1 \leq m \leq 3. \qquad (1)$$

This is a system of three equations with respect to the unknown parameters $A(P)$, $B(P)$ and $C(P)$. Its solution writes classically:

$$A(P) = \frac{I_1(P) + I_2(P) + I_3(P)}{3}, \qquad B(P) = \frac{2I_1(P) - I_2(P) - I_3(P)}{3} \quad \text{and} \quad C(P) = \frac{I_2(P) - I_3(P)}{\sqrt{3}}. \qquad (2)$$

Regular use of these relations is made in the following sections. The reader may also refer to the nomenclature of Appendix A for a complete list of employed mathematical notations.

**3.2 Phase-shifting in telescope pupil plane**

This case is very similar to the PSTI original scheme [12], to which a coronagraphic mask is added into the image plane. All the mathematical development is illustrated in Figure 3, including the expressions of complex amplitudes, measured intensities and WFE reconstruction process. The complex amplitude in the OXY pupil plane is firstly written as:

$$A_m(P) = B_d^D(P)\exp[ikW(P)] + B_d(P)\exp[i\phi_m], \qquad (3)$$

with $k = 2\pi/\lambda$ the wavenumber of the electromagnetic radiation, $B_d^D(P)$ and $B_d(P)$ the transmission maps of the main and reference pupils as defined in Appendix A, $W(P)$ the WFE to be measured and $\phi_m$ the $m^{th}$ phase-shift. Following

Fraunhofer diffraction theory, he complex amplitude in the O'X'Y' image plane is obtained via the Fourier transform of $A_m(P)$:

$$\hat{A}'_m(M') = \hat{B}^D_d(M') * FT[\exp[ikW(P)]](M') + \hat{B}_d(M')\exp[i\phi_m], \quad (4)$$

where $FT[\ ]$ and $*$ denote Fourier transform and convolution operation respectively. Assuming that the coronagraph is a PMC with phase function denoted $\varphi(M')$, the transmitted wave is expressed as;

$$\hat{A}'_m(M') = \exp[i\varphi(M')] \times \left(\hat{B}^D_d(M') * FT[\exp[ikW(P)]](M') + \hat{B}_d(M')\exp[i\phi_m]\right). \quad (5)$$

After inverse Fourier transform, the complex amplitude in the Lyot stop plane O"X"Y" writes as:

$$A''_m(P'') = \left[B^D_d(P'')\exp[ikW(P'')]\right] * FT^{-1}(\exp[i\varphi(M')])(P'') + \exp[i\phi_m]B_d(P'') * FT^{-1}(\exp[i\varphi(M')])(P''), \quad (6)$$

and after passing through the Lyot stop defined by function $B_{D_L}(P'')$:

$$A''_m(P'') = B_{D_L}(P'')\left(\left[B^D_d(P'')\exp[ikW(P'')]\right] * FT^{-1}(\exp[i\varphi(M')])(P'')\right) \\ + \exp[i\phi_m]B_{D_L}(P'')\left(B_d(P'') * FT^{-1}(\exp[i\varphi(M')])(P'')\right). \quad (7)$$

One last direct Fourier transform gives access to the wave amplitude focused at the coronagraphic plane O'''X'''Y''':

$$A'''_m(M''') = \hat{B}_{D_L}(M''') * \left(\exp[i\varphi(M''')] \times \left[\hat{B}^D_d(M''') * FT(\exp[ikW(P'')])(M''')\right]\right) \\ + \exp[i\phi_m]\hat{B}_{D_L}(M''') * \left(\hat{B}_d(M''')\exp[i\varphi(M''')]\right). \quad (8)$$

In this phase-shifting scheme, intensity variations are sensed into that last image plane (see Figure 2-2). After multiplying $A'''_m(M''')$ with its complex conjugate and rearranging the terms, they can be written as:

$$I'''_m(M''') = |A'''_m(M''')|^2 = |\hat{O}(M''')|^2 + |\hat{O}_R(M''')|^2 + 2\cos\phi_m \operatorname{Re}[\hat{O}(M''')\hat{O}^*_R(M''')] + 2\sin\phi_m \operatorname{Im}[\hat{O}(M''')\hat{O}^*_R(M''')], \quad (9a)$$

with:
$$\hat{O}(M''') = \hat{B}_{D_L}(M''') * \left[\exp[i\varphi(M''')] \times \left[\hat{B}^D_d(M''') * FT(\exp[ikW(P'')])(M''')\right]\right] \quad (9b)$$
$$\hat{O}_R(M''') = \hat{B}_{D_L}(M''') * \left(\hat{B}_d(M''')\exp[i\varphi(M''')]\right), \quad (9c)$$

and Re[ ] and Im[ ] respectively stand for the real and imaginary part of a complex number. It must be noted that the searched WFE $W(P)$ actually is identical to the term $W(P'')$ in the right hand side of Eq. 9b, since there is a magnification factor of one between the telescope exit pupil and Lyot stop planes.

Eqs. 9 demonstrate that the wavefront error could only be reconstructed at the price of a deconvolution process, which is not desirable for AO application. However the extraction process can be dramatically simplified by use of the two following hypotheses, named "Delta" approximations in pupil and image planes.

**A. Delta approximation in pupil plane**

The principle and justification for that first Delta approximation were extensively discussed in Refs. [7] and [12]. To summarize, it is assumed that the diameter $d$ of the reference pupil is so small with respect to the diameter $D$ of the main pupil (in other words $d/D \ll 1$), so that the reference pupil function $B_d(P'')$ can be approximated by $\rho^2\delta(P'')$, where $\delta(P'')$ is the Dirac "Delta" function. Hence $\hat{B}_d(M''') = \rho^2$ and Eq. 9c simplifies as:

$$\hat{P}_R(M''') \approx \rho^2 \hat{B}_{D_L}(M''') * \exp[i\varphi(M''')], \quad (10)$$

**B. Delta approximation in image plane**

This is indeed an inverted version of the previous approximation that is here applied to the Lyot stop. It states that if the stop size is large enough (i.e. $D_L \gg D$), then $\hat{B}_{D_L}(M''')$ can be approximated to the Dirac function $\delta(M''')$ in relations 9b and 10, yielding:

$$\hat{O}(M''') \approx \exp[i\varphi(M''')] \times \left[\hat{B}_d^D(M''') * FT(\exp[ikW(P'')])(M''')\right] \quad (11a)$$

and: $\quad \hat{O}_R(M''') \approx \rho^2 \exp[i\varphi(M''')], \quad (11b)$

Then the analytical expressions of the measured intensities in the coronagraphic plane are again simplified as:

$$I_m'''(M''') \approx |\hat{P}(M''')|^2 + \rho^4 + 2\rho^2 \cos\phi_m \, \text{Re}[\hat{P}(M''')] + 2\rho^2 \sin\phi_m \, \text{Im}[\hat{P}(M''')], \quad (12a)$$

where: $\quad \hat{P}(M''') = \hat{B}_d^D(M''') * FT(\exp[ikW(P'')])(M'''). \quad (12b)$

However this last approximation has an important consequence: since an oversized Lyot stop is of no interest to coronagraphy, it follows that the WFS arm shall be separated from the science detector arm before the real Lyot stop of the coronagraph, for example by means of a dichroic plate. WFE measurement shall then be carried out in an independent image plane optically conjugated with the coronagraphic plane and without Lyot stop. This point is further discussed in section 4 and illustrated by the proposed optical design of section 5.

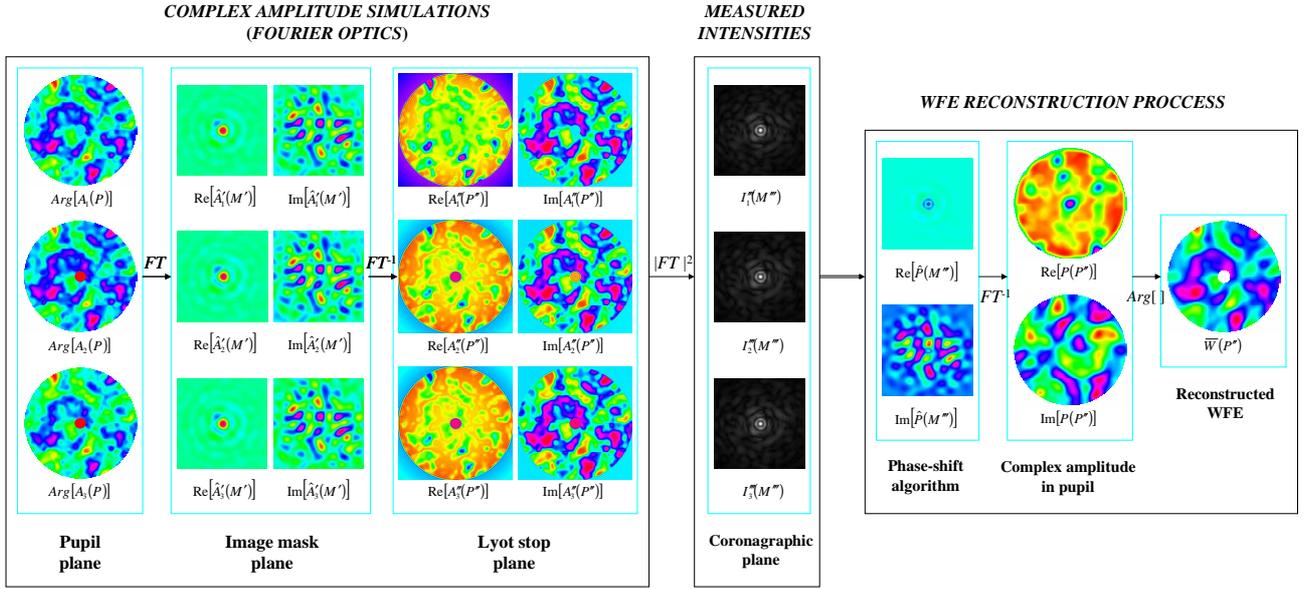

Figure 3: Schematic view of complex amplitudes propagation (left side), measured intensities (centre) and WFE reconstruction process (right side) in pupil plane phase-shifting mode. For illustration the coronagraph is equipped with a Roddier phase mask. Phase, real part and imaginary part of complex functions are displayed in false colors and their intensity in gray scale. Mathematical notations are the same as those employed in the text and in Appendix A.

**C. WFE reconstruction**

From the three measured intensities $I_1'''(M''')$, $I_2'''(M''')$ and $I_3'''(M''')$, the phase-shift algorithm of relations 1-2 allows calculating the two real functions $B(P'')$ and $C(P'')$ as:

$$B(M''') = \frac{2I_1''(M''') - I_2''(M''') - I_3''(M''')}{3} = 2\rho^2 \, \text{Re}[\hat{P}(M''')] \quad (13a)$$

$$C(M''') = \frac{I_2''(M''') - I_3''(M''')}{\sqrt{3}} = 2\rho^2 \, \text{Im}[\hat{P}(M''')], \quad (13b)$$

from which the complex function $\hat{P}(M''')$ can be reconstructed. Noting from Eq. 12b that its inverse Fourier transform $P(P'')$ is equal to $\hat{B}_d^D(P'')\exp[ikW(P'')]$, the searched WFE is finally reconstructed as:

$$\overline{W}(P'') = \frac{\lambda}{2\pi} Arg[P(P'')]. \qquad (14)$$

where *Arg*[ ] denotes the real argument of a complex number.

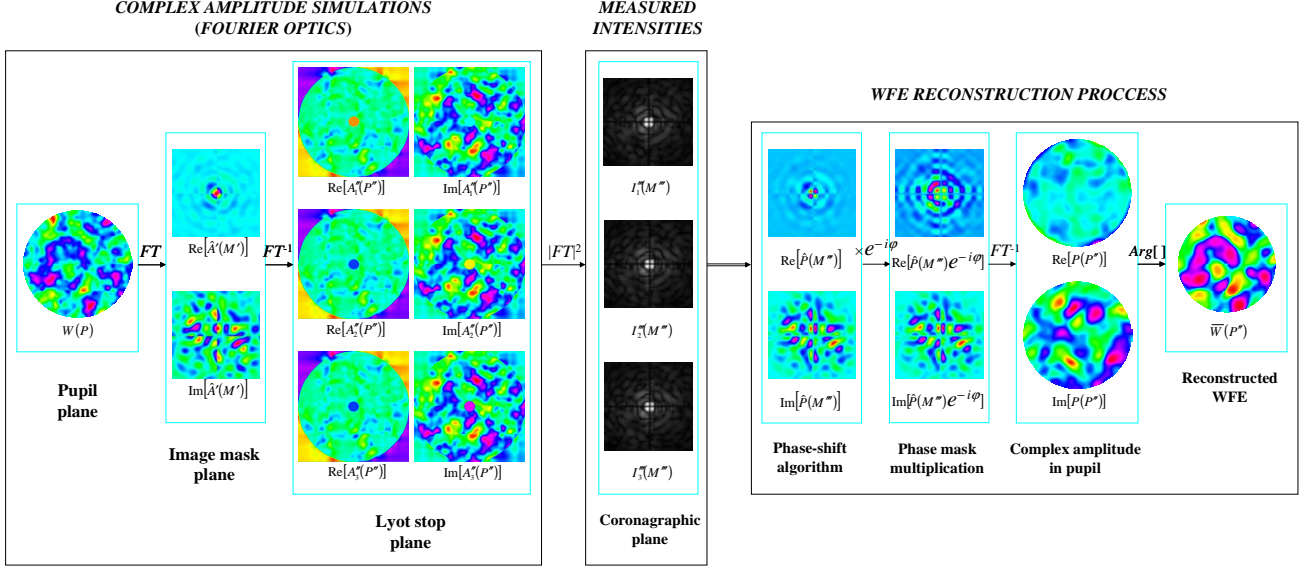

Figure 4: Same illustrations as in Figure 3 with phase-shift introduction into the Lyot stop plane and a Four-quadrant phase-mask coronagraph.

## 3.3 Phase-shifting in Lyot stop plane

From mathematics point of view, this is a variant of the previous case where phase-shifts are introduced into the Lyot stop plane (at point O" in Figure 2-2) or one of its optical conjugate. Following the logic illustrated in Figure 4 and applying both Delta approximations in § 3.2.A and 3.2.B, Eqs. 6, 9a and 10a become respectively:

- Complex amplitude after phase mask:

$$\hat{A}'(M') = \left(\hat{B}_d^D(M') * FT[\exp[ikW(P)]](M')\right) \times \exp[i\varphi(M')]. \qquad (15)$$

- Complex amplitude after Lyot stop:

$$A''_m(P'') = \left[B_d^D(P'')\exp[ikW(P'')]\right] * FT^{-1}(\exp[i\varphi(M')])(P'') + \rho^2 \exp[i\phi_m]\delta(P'') \qquad (16)$$

- Measured intensities in coronagraphic plane:

$$I'''_m(M''') = \left|\hat{P}(M''')\right|^2 + \rho^4 + 2\rho^2 \cos\phi_m \operatorname{Re}[\hat{P}(M''')] + 2\rho^2 \sin\phi_m \operatorname{Im}[\hat{P}(M''')], \qquad (17a)$$

$$\text{where:} \quad \hat{P}(M''') = \exp[i\varphi(M''')] \times \left[\hat{B}_d^D(M''') * FT(\exp[ikW(P'')])(M''')\right]. \qquad (17b)$$

It can be seen that Eq. 17a is analogous to Eq. 12a, but the presence of a multiplying term $\exp[i\varphi(M''')]$ into the expression of $\hat{P}(M''')$ involves a slight modification of the WFE measurement procedure. Assuming that the coronagraph phase function $\varphi(M''')$ is known, either by design o after characterization tests, the complex function $P(P'')$ can be built by multiplying $\hat{P}(M''')$ with the complex conjugate of the phase mask, then computing the inverse Fourier transform of the result:

$$P(P'') = FT^{-1}\left(\hat{P}(M''')\exp[-i\varphi(M''')]\right)(P'') = \hat{B}_d^D(P'')\exp[ikW(P'')]. \quad (18)$$

The WFE is finally reconstructed by using Eq. 14 as in the previous sub-section.

### 3.4 Phase-shifting in coronagraphic image plane

In this sub-section is finally studied the case when phase-shifts are introduced into an image plane located behind the phase mask. Measurements are performed in an optical conjugate of the telescope pupil plane OXY as illustrated in Figure 2-3. The whole mathematical development is illustrated in Figure 5 and the employed notations are summarized in the Appendix A.

Let us first write the complex amplitude after the phase mask as:

$$\hat{A}'(M') = \exp[i\varphi(M')] \times \left[\hat{B}_D(M') * TF(\exp[ikW(P)])(M')\right]. \quad (19)$$

Here again it is assumed that the Delta approximation in the image plane is applicable ( of § 3.2.B), and that the WFE is sensed from a parallel optical arm without Lyot stop. Hence the complex amplitude after phase-shift introduction can be expressed as:

$$\hat{A}'_m(M') = [1 - B_{d'}(M')]\exp[i\varphi(M')] \times \left[\hat{B}_D(M') * FT(\exp[ikW(P)])(M')\right] \\ + \exp[i\phi_m]B_{d'}(M')\exp[i\varphi(M')] \times \left[\hat{B}_D(M') * FT(\exp[ikW(P)])(M')\right] \quad (20)$$

where function $B_{d'}(M')$ is defining the circular phase-shifting area into the O'X'Y' plane (see Appendix A). The absence of Lyot stop also gives the opportunity to simplify coordinates notation, using 'superscripts for points M' located in any image plane and "superscripts for points P" located into the pupil planes. Then the complex amplitude in the measurement plane is obtained by inverse Fourier transforming $\hat{A}'_m(M')$, which leads to:

$$A''_m(P'') = FT^{-1}(\exp[i\varphi(M')])(P'') * [B_D(P'')\exp[ikW(P'')]] \\ + (\exp[i\phi_m] - 1)\hat{B}_{d'}(P'') * FT^{-1}(\exp[i\varphi(M')])(P'') * [B_D(P'')\exp[ikW(P'')]] \quad (21)$$

### A. Delta approximation in the image plane

Here also this Delta approximation may be used, tough under a slightly different manner: assuming that $B_{d'}(M')$ can be replaced with a Dirac function $\delta(M')$, it follows that its Fourier transform $\hat{B}_{d'}(P'')$ can be approximated as a uniform function equal to unity. Therefore the triple convolution product in Eq. 21 tends toward a fixed complex number noted $R_0 \exp[i\phi_0]$ with $R_0$ its modulus and $\phi_0$ its phase. Thus $A''_m(P'')$ reduces to:

$$A''_m(P'') = FT^{-1}(\exp[i\varphi(M')])(P'') * [B_D(P'')\exp[ikW(P'')]] + R_0 \exp[i\phi_0](\exp[i\phi_m] - 1). \quad (22)$$

The basic condition for this Delta approximation to be valid is that the diameter $d'$ of the phase-shifting zone is small with respect to the first lobe of the non aberrated PSF in the image plane, i.e. $d' \ll \lambda F/D$. After multiplying $A''_m(P'')$ with its complex conjugate and rearranging the terms, the intensity variations $I''_m(P'')$ sensed into the pupil plane write as:

$$I''_m(P'') = |A''_m(P'')|^2 = |P(P'')|^2 - 2R_0 \cos\phi_0 \operatorname{Re}[P(P'')] - 2R_0 \sin\phi_0 \operatorname{Im}[P(P'')] + 2R_0^2 - 2R_0^2 \cos\phi_m \\ + 2R_0 \cos(\phi_m + \phi_0)\operatorname{Re}[P(P'')] + 2R_0 \sin(\phi_m + \phi_0)\operatorname{Im}[P(P'')] \quad (23a)$$

with: $\quad P(P'') = FT^{-1}(\exp[i\varphi(M')])(P'') * [B_D(P'')\exp[ikW(P'')]]. \quad (23b)$

Because the sole effect of the constant phase $\phi_0$ is to add an arbitrary piston to the searched WFE term $W(P'')$, it can be set to zero, thus allowing to simplify Eq. 23a:

$$I''_m(P'') = |P(P'')|^2 - 2R_0 \, \text{Re}[P(P'')] + 2R_0^2 + 2R_0 \cos\phi_m \left(\text{Re}[P(P'')] - R_0\right) + 2R_0 \sin\phi_m \, \text{Im}[P(P'')]. \quad (24)$$

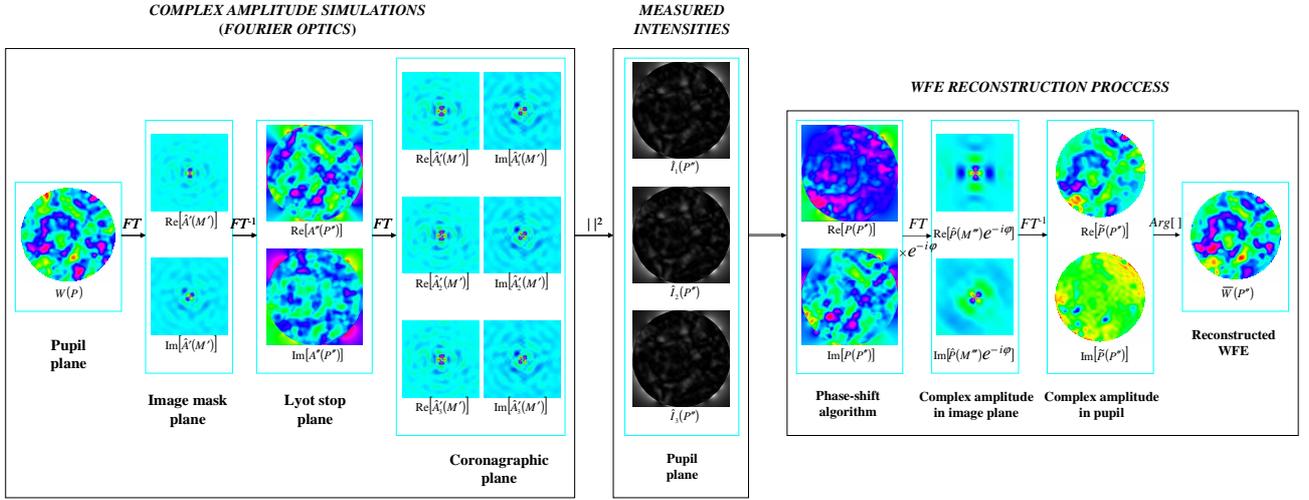

Figure 5: Same illustrations as in Figure 3 and 4 with phase-shift introduction into the coronagraphic image plane and a vortex coronagraph.

**B. WFE reconstruction**

From the three measured intensities in the pupil plane $I''_1(P'')$, $I''_2(P'')$ and $I''_3(P'')$, the phase-shift algorithm of relations 1-2 allows determining the three real functions $A(P'')$, $B(P'')$ and $C(P'')$:

$$A(P'') = \frac{I''_1(P'') + I''_2(P'') + I''_3(P'')}{3} = |P(P'')|^2 - 2R_0 \, \text{Re}[P(P'')] + 2R_0^2 \quad (25a)$$

$$B(P'') = \frac{2I''_1(P'') - I''_2(P'') - I''_3(P'')}{3} = 2R_0 \left(\text{Re}[P(P'')] - R_0\right) \quad (25b)$$

$$C(P'') = \frac{I''_2(P'') - I''_3(P'')}{\sqrt{3}} = 2R_0 \, \text{Im}[P(P'')]. \quad (25c)$$

At all points P" into the pupil plane, the relations 25 form a system of three equations to the three unknown $R_0$, $\text{Re}[P(P'')]$ and $\text{Im}[P(P'')]$. $R_0$ is firstly determined noting that Eqs. 25 give two independent expressions of $|O(P'')|^2$, leading to a second-order equation vs. $R_0^2$:

$$R_0^4 - R_0^2 A(P'') + \frac{B^2(P'') + C^2(P'')}{4} = 0, \quad (26a)$$

whose valid solution is:
$$R_0^2(P'') = \frac{A(P'') - \sqrt{A^2(P'') - B^2(P'') - C^2(P'')}}{2}. \quad (26b)$$

Practically $R_0$ can be calculated as $\sqrt{\langle R_0^2(P'') \rangle}$, where brackets stand for spatial averaging over the pupil area. The next step is then to build the complex function $P(P'')$ as:

$$P(P'') = \frac{B(P'')}{2R_0} + R_0 + i\frac{C(P'')}{2R_0}. \quad (27)$$

From Eq. 23b the Fourier transform of $P(P'')$ is found to be:

$$\hat{P}(M''') = FT(P(P''))(M''') = \exp[i\varphi(M''')] \times [FT(B_D(P'')\exp[ikW(P'')])(M''')], \quad (28)$$

and after multiplying $\hat{P}(M')$ with the complex conjugate of the PMC wave function, a last inverse Fourier transform allows extracting the searched WFE:

$$\overline{W}(P'') = \frac{\lambda}{2\pi} Arg[\tilde{P}(P'')], \quad (29a)$$

with: $\tilde{P}(P'') = FT^{-1}(\hat{P}(M''')\exp[-i\varphi(M''')])(P'')$. (29b)

Finally, it should be noted that a similar mathematical development can be developed in the case when phase-shifts would be introduced before the phase mask of the coronagraph, i.e. into an intermediate image plane of the telescope. This development actually leads to the same analytical formulas and measurement procedure.

## 4 NUMERICAL SIMULATIONS

Extensive numerical simulations have been carried out in order to compare the measurement accuracy of the three phase-shifting processes described in the section 3, and to identify their critical system parameters. In this section are presented all the simulated cases (§ 4.1), the main achieved results and their consequence on the choice of a phase-shifting plane (§ 4.2), and the influence of the Lyot stop diameter on achievable performance (§ 4.3).

### 4.1 Numerical model and studied cases

The three phase-shifting schemes considered in this study are summarized in Table 1. Their numerical models strictly follow the logic described in sections 3.2 to 3.4 and illustrated in the Figure 3 to 5 correspondingly. They are all separated into two independent modules, respectively shown on the left and right sides of the Figures. The computer programs are written in IDL language.

- The first module is common to all phase-shifting cases. It is basically a Fourier optics model propagating complex amplitudes throughout the optical system from the input WFE function $W(P)$ to the measurement plane via Fourier transforms, and multiplying them with the complex transmission functions of the PMC. Only Franhofer diffraction is taken into account, Fresnel diffraction being neglected. The intensities are taken as the square modulus of the final complex amplitudes and stored into external files (one for each phase-shift $\phi_1$, $\phi_2$ and $\phi_3$). The sequence of employed algorithms is summarized in Figure 6. The functions $W(P)$ were typically sampled by 512 x 512 points and embedded into 4096 x 4096 arrays in order to prevent possible aliasing generated by fast Fourier transform (FFT) algorithm.

- The second module reconstructs the entrance WFE from the measured intensities. Two variants have been developed, depending whether phase-shifts are introduced into a pupil or image plane. Their algorithmic sequences are also given in Figure 6.

We take the example of a large-size coronagraph telescope of 10-m diameter and focal length $F$ = 200 m, thus having an aperture number $F/D$ = 20. All simulations are monochromatic and performed at the wavelength $\lambda$ = 0.5 µm. The telescope is equipped with a PMC whose typical phase functions may be of three different types:

- A Roddier phase mask (RPM) made of a small central circular area with amplitude transmission factor of –1 [20], such that the phase function is defined as:

$\varphi(M') = \pi$     if $0 \leq \rho' \leq 0.4842\ \lambda F/D$    and: (30)
$\varphi(M') = 0$     elsewhere,

with $\rho'$ the radial coordinate of point M' in the image plane (see Figure 2). This phase mask achieves perfect extinction at the Lyot stop centre when it satisfies to the condition $J_0(\pi D\rho'/\lambda F) = 0.5$, where $J_0$ is the type-J Bessel function at the order zero.

- A Four-quadrant phase-mask (FQPM) such as originally described in Ref. [21], made of four identical angular sections, each one being phase-shifted by π with respect to its neighbours:

$$\varphi(M') = 0 \quad \text{if} \quad 0 \leq \theta' < \pi/2 \quad \text{and} \quad \pi \leq \theta' < 3\pi/2 \quad (31)$$
$$\varphi(M') = \pi \quad \text{if} \quad \pi/2 \leq \theta' < \pi \quad \text{and} \quad 3\pi/2 \leq \theta' < 2\pi,$$

with θ' the polar angle of point M' (see Figure 2).

- A Vortex phase-mask (VPM) of topological charge $m = 2$ where the phase function is varying linearly from 0 to 4π as function of the polar angle:

$$\varphi(M') = 2\theta'. \quad (32)$$

A general theory of VPM coronagraphs can be found in Ref. [22].

In Figure 3 to Figure 5 are illustrated the effects of these three different types of PMC on complex amplitudes propagated throughout the optical system. In addition and for each phase-shifting process and PMC family, two typical cases of entrance WFE $W(P)$ were considered. They are both represented in the left side of Figure 7 and essentially differ by their spatial frequency content:

o A low spatial frequency WFE standing for non common optical aberrations (NCPA) arising between the XAO and science detectors. Those NCPA usually are quasi-static and should be characterized by a specific WFS located inside the coronagraph. The function $W(P)$ was generated from the first 16 Zernike polynomials whose amplitudes are selected randomly. It was then rescaled to have a Peak-to-Valley (PTV) value around 0.5λ, in order to evaluate the measurement accuracy of the phase-shifting process outside of the weak aberrations domain.

o A mid-frequency WFE representing possible XAO residuals, i.e. rapidly varying atmospheric disturbance that was not totally cancelled by the XAO deformable mirror. The function $W(P)$ was built from random amplitude cells of average width $D/20$. Here again the amplitude of the WFE was adjusted to be ~ 0.5λ PTV, even if those residuals are expected to be much smaller for real AO systems.

For the three considered phase-shifting schemes, numerical simulations using the here above coronagraph phase functions $\varphi(M')$ and wavefront errors $W(P)$ have been carried out, and their results are reported in the next sub-section.

Table 1: Summary of numerical models.

| Phase-shifting plane | Measured intensities | Procedure description | Illustration |
|---|---|---|---|
| Telescope pupil plane | In coronagraphic plane | Section 3.2 | Figure 3 (with RPM) |
| Lyot stop plane | In coronagraphic plane | Section 3.3 | Figure 4 (with FQPM) |
| Coronagraphic plane | In pupil plane | Section 3.4 | Figure 5 (with VPM) |

## 4.2 Choice of phase-shifting plane

The main results of the numerical models are summarized in Table 2 and Table 3, for the low and mid spatial frequency WFE functions $W(P)$ respectively. The tables are divided into three main columns for each phase-shifting process. In those columns are indicated the RMS and PTV values of the reconstructed WFE $\overline{W}(P'')$, of the difference maps between original and estimated wavefronts $\Delta W(P'') = \overline{W}(P'') - W(P)$, and of the relative measurement errors $|\Delta W(P'')|/W(P)$ expressed in percent of the original WFE. In each row is indicated the coronagraph type, starting with a reference case without phase mask in the image plane to serve as comparison basis. For each phase-shifting case an horizontal strip specifies the main simulation parameters. They are:

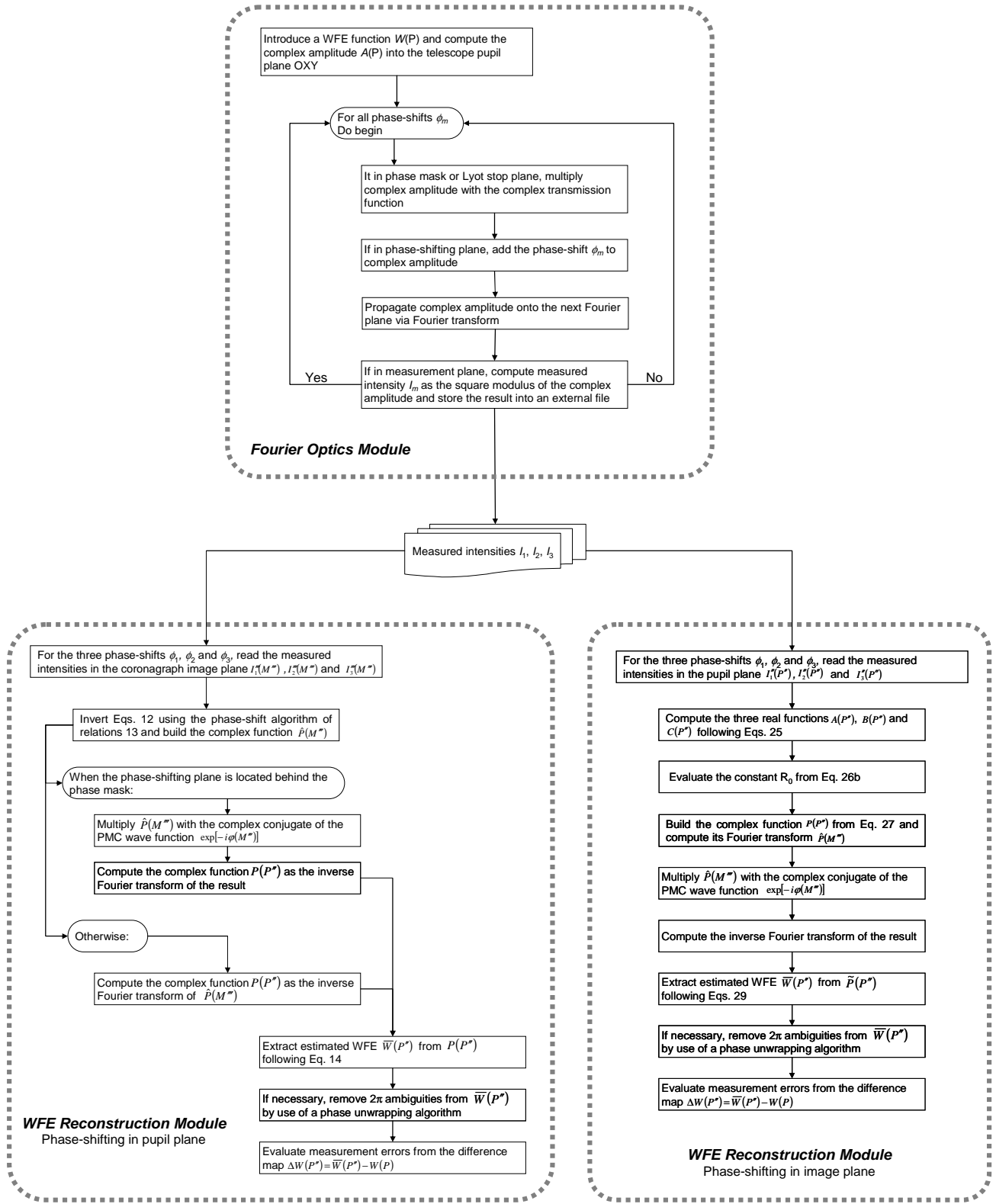

Figure 6: Flow-chart of the numerical models simulating the measured intensities and reconstructing the entrance WFE.

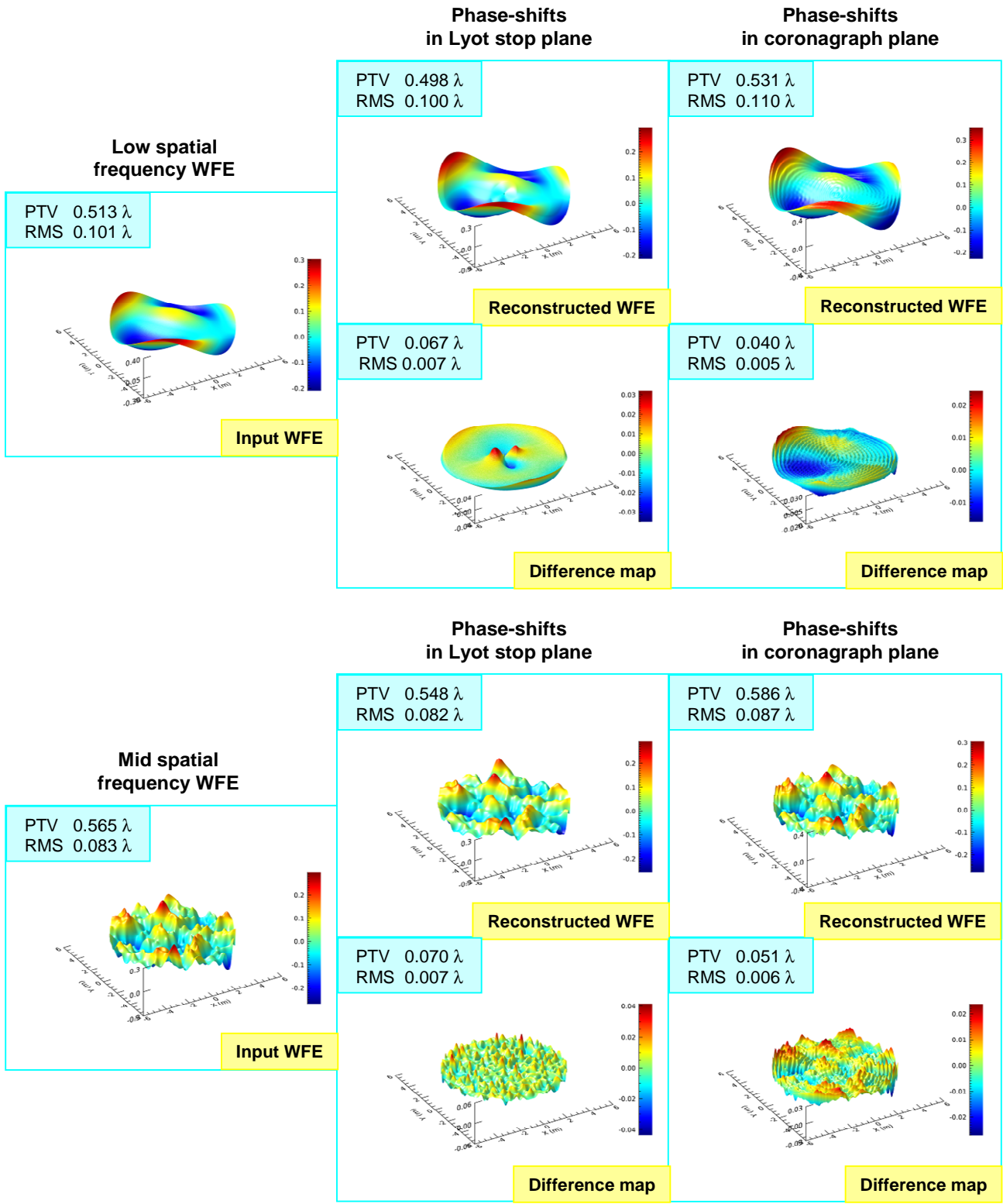

Figure 7: Surface plots of the input and reconstructed WFE functions. Left side: entrance wavefronts. Centre: reconstructed wavefronts and their difference maps when phase-shifts are introduced into the Lyot stop plane. Right side: same plots for phase-shifting in the image plane.

- The ratio $\rho = d/D$ between the diameters of the reference and main pupils when phase-shifts are introduced in a pupil plane (either telescope exit pupil or Lyot stop). It is actually the prime marker for the first Delta approximation of section 3.2.A. From Ref. [9] $\rho = 0.05$ is a good compromise for sensing low frequency WFEs. It had however to be decreased to $\rho = 0.02$ in the mid-frequency case.

- The ratio $\Lambda = D_L/D$ between the diameters of the Lyot stop and telescope entrance pupil, which is closely related to the second Delta approximation of section 3.2.B: assuming function $\hat{B}_{D_L}(M''')$ to be a Dirac distribution theoretically implies a Lyot stop of infinite size, hence the higher $\Lambda$, the more valid the approximation. Practically the parameter was set to $\Lambda = 4$ for all simulations presented in this sub-section.

- The ratio $\eta = \overline{D}/D$ defining the useful pupil diameter $\overline{D}$ where reconstructed wavefronts $\overline{W}(P'')$ are compared with the input WFE $W(P)$. This parameter was introduced because reconstructed WFEs often exhibit larger errors near the pupil rim (typically for $0.95 \leq \eta \leq 1$) that were not taken into account for evaluating the measurement accuracy. However, identical $\eta$ parameters are always used for a given type of coronagraph or entrance WFE.

- The diameter $d'$ of the phase-shifting area when phase-shifts are introduced in an image plane. Normalizing $d'$ with the nominal diameter of the Airy disk $2.44\lambda F/D$ allows defining the ratio $\varepsilon = 0.41\rho'D/\lambda F$ expressed in fraction of the central core of the non aberrated PSF. Practically, this parameter has been set to $\varepsilon = 0.1$ in all simulated cases, excepting the RPM coronagraph for which better measurement accuracy is achieved when $\varepsilon = 0.6$.

Numerical simulations are also illustrated in Figure 7 for a vortex coronagraph equipped with a VPM of charge 2. For both the low and mid-frequency cases, it shows surface plots of the entrance wavefront (left side of figure), the reconstructed wavefront $\overline{W}(P'')$ and its difference map $\Delta W(P'')$ with phase-shifting introduction into the Lyot stop plane (central part of figure), and the same plots when introducing phase-shifts in the image plane (right side of figure). All the information presented in the figure and in Table 2 and 3 is quite rich and is summarized below, not necessarily by order of importance but following the columns of the tables.

a. Starting with the reference case where there is no coronagraph and phase-shifts are introduced into the telescope pupil plane OXY, it can be verified that the relative measurement accuracy lays in the 5-10 % range in RMS sense (or equivalently around $\lambda/150$). The best accuracy is achieved for the low spatial frequency WFE. This result is in good agreement with Refs. [9] and [12] where the case was studied originally. Not surprisingly, the same numbers are obtained when phase-shifts are introduced into the Lyot stop plane O"X"Y".

b. The achieved measurement accuracy is even better when phase-shifts are introduced into the coronagraphic image plane, since relative errors lay in the 2-4 % range. Moreover, it offers wider measurement areas (characterized by parameter $\eta$), eliminating for example surface indetermination near the pupil centre that is inherent to the pupil plane phase-shifting process. It may thus be concluded that phase-shifting in the image plane is superior to phase-shifting in the pupil plane, at least when no coronagraph is involved. These results also are consistent with those reported in Ref. [17], without being limited to the weak aberration domain nor to a few Zernike modes.

c. Considering now the case of coronagraphs, it can be seen that the measurement errors are not significantly degraded, either when phase-shifts are introduced into a pupil or image plane. Relative errors are ranging from 5 to 10 % for the low frequency WFE and from 10 to 15 % for the mid frequency WFE. Therefore phase-shifting methods look well-suited to wavefront sensing from behind a coronagraph mask.

d. A few distinctions shall however be made between the type of phase masks: the best accuracy seems to be attained with the RPM and VPM (they are equivalent to the case without coronagraph), while it is degraded by a few percents when using a FQPM. This may be due to the excessive dispersing power of that phase mask, eventually requiring Lyot stop diameters larger than $D_L = 4D$, i.e. $\Lambda > 4$.

e. Finally, and despite small variations between them, it is found that the measurement errors of the low and mid-frequency WFE are of the same magnitude order, whatever the type of coronagraph or phase-shift location. It can be concluded that the phase-shifting processes are not limited to low order wavefront sensing, nor to the weak aberrations domain.

Table 2: Measurement accuracy of the low spatial frequency WFE. The input RMS and PTV values are 0.101 λ and 0.513 λ respectively.

| Type of Coronagraph | PHASE-SHIFT LOCATION | | | | | | | | | |
|---|---|---|---|---|---|---|---|---|---|---|
| | Telescope pupil plane | | | Lyot stop plane | | | Image plane | | | |
| | $\rho = 0.05$; $\Lambda = 4$; $0.1 < \eta < 0.9$ | | | $\rho = 0.05$; $\Lambda = 4$; $0.1 < \eta < 0.9$ | | | $\varepsilon = 0.1^{(*)}$; $\Lambda = 4$; $0 < \eta < 0.95$ | | | |
| | Measured (waves) | Difference (waves) | Relative error (%) | Measured (waves) | Difference (waves) | Relative error (%) | Measured (waves) | Difference (waves) | Relative error (%) | |
| No coronagrah | 0.098 | 0.005 | 5 | 0.098 | 0.005 | 5 | 0.106 | 0.002 | 2 | RMS |
| | 0.489 | 0.037 | 7 | 0.489 | 0.036 | 7 | 0.516 | 0.017 | 3 | PTV |
| Roddier | 0.107 | 0.008 | 8 | 0.110 | 0.009 | 9 | 0.104 | 0.007 | 6 | RMS |
| | 0.520 | 0.072 | 14 | 0.540 | 0.035 | 7 | 0.504 | 0.039 | 8 | PTV |
| 4-Quadrants | 0.108 | 0.009 | 9 | 0.107 | 0.008 | 8 | 0.114 | 0.022 | 21 | RMS |
| | 0.519 | 0.079 | 15 | 0.518 | 0.046 | 9 | 0.542 | 0.080 | 16 | PTV |
| Vortex (m=2) | 0.101 | 0.004 | 4 | 0.100 | 0.007 | 7 | 0.110 | 0.006 | 5 | RMS |
| | 0.502 | 0.042 | 8 | 0.498 | 0.067 | 13 | 0.531 | 0.040 | 8 | PTV |

(*) ε = 0.6 for Roddier phase mask

Table 3: Measurement accuracy of the mid spatial frequency WFE. The input RMS and PTV values are 0.083 λ and 0.565 λ respectively.

| Type of Coronagraph | PHASE-SHIFT LOCATION | | | | | | | | | |
|---|---|---|---|---|---|---|---|---|---|---|
| | Telescope pupil plane | | | Lyot stop plane | | | Image plane | | | |
| | $\rho = 0.02$; $\Lambda = 4$; $0.04 < \eta < 0.95$ | | | $\rho = 0.02$; $\Lambda = 4$; $0.04 < \eta < 0.95$ | | | $\varepsilon = 0.1^{(*)}$; $\Lambda = 4$; $0 < \eta < 0.95$ | | | |
| | Measured (waves) | Difference (waves) | Relative error (%) | Measured (waves) | Difference (waves) | Relative error (%) | Measured (waves) | Difference (waves) | Relative error (%) | |
| No coronagrah | 0.078 | 0.008 | 10 | 0.078 | 0.008 | 10 | 0.083 | 0.003 | 4 | RMS |
| | 0.534 | 0.082 | 14 | 0.534 | 0.080 | 14 | 0.567 | 0.028 | 5 | PTV |
| Roddier | 0.088 | 0.009 | 11 | 0.091 | 0.009 | 10 | 0.083 | 0.005 | 7 | RMS |
| | 0.565 | 0.074 | 13 | 0.592 | 0.041 | 7 | 0.544 | 0.044 | 8 | PTV |
| 4-Quadrants | 0.089 | 0.010 | 12 | 0.089 | 0.010 | 12 | 0.096 | 0.025 | 30 | RMS |
| | 0.572 | 0.080 | 14 | 0.569 | 0.067 | 12 | 0.594 | 0.109 | 19 | PTV |
| Vortex (m=2) | 0.082 | 0.007 | 9 | 0.082 | 0.007 | 9 | 0.087 | 0.006 | 7 | RMS |
| | 0.550 | 0.071 | 13 | 0.548 | 0.070 | 12 | 0.586 | 0.051 | 9 | PTV |

(*) ε = 0.6 for Roddier phase mask

## 4.3 Effect of Lyot stop

In this section is discussed the validity of the so-called "Delta approximation in the image plane" of section 3.2.B and 3.4.A. This approximation implies that the Lyot stop is large enough to replace the Fourier transform $\hat{B}_{D_L}(M''')$ of its transmission map with the Dirac distribution. It follows that ideally the wavefront sensing optics should be separated from the science arm, and only the latter should incorporate a Lyot stop. For practical reasons however the WFS optics could not be of infinite size, hence the need for evaluating the achievable measurement accuracy as function of the pupil diameter ratio $\Lambda = D_L/D$. This is the goal of the following numerical simulations.

Based on the results of the previous section, we selected the typical case of a low frequency input WFE entering a vortex coronagraph of topological charge $m = 2$. For both pupil and image plane phase-shifting modes, the achieved measurement accuracy in terms of RMS and PTV values of the relative error $|\Delta W(P'')|/W(P)$ is estimated in Table 4 at different Λ ratios. The results are also illustrated in Figure 8, showing sensitivity curves in the range $1 \leq \Lambda \leq 8$ and surface plots of two typical error maps $\Delta W(P'')$. The curves in Figure 8-A demonstrate that a good measurement accuracy (within 7-10% in RMS sense) is achieved when $\Lambda \geq 3$ whatever the phase-shifting process. This criterion may thus be considered as a basic requirement for the wavefront sensing optics. Below $\Lambda = 3$ the measurement accuracy progressively deteriorates. The term "Fail" in Table 4 when $\Lambda = 1$ indicates that the extracted WFE could not be unwrapped used simple algorithms, thus showing residual $2\pi$-jumps which may rule out this case for AO application. We also took the opportunity to study the case of a "negative" Lyot stop, meaning that its transmission function in Eq. 7 is defined as $1 - B_{D_L}(P'')$ instead of $B_{D_L}(P'')$. Practically, this case has been tested for low order wavefront sensing using light reflected by the Lyot stop [23]. Results indicated in the first row of Table 4 and in Figure 8-C seem however rather disappointing, showing that only a small part of the low spatial frequency perturbation could be restored. Hence this method is probably limited to closed-loop operation on low order Zernike modes, and not applicable to phase-shifting technique.

Table 4: Effect of Lyot stop size on wavefront sensing accuracy (case of low order WFE and vortex coronagraph).

| $D_L/D$ ratio | Relative errors (%) | | | |
| --- | --- | --- | --- | --- |
| | Phase-shift in Lyot stop plane | | Phase-shift in coronagraph plane | |
| | RMS | PTV | RMS | PTV |
| *Negative* | 89 | 85 | 85 | 85 |
| 1 | *Fail* | | *Fail* | |
| 1.5 | 34 | 23 | 37 | 35 |
| 2 | 17 | 16 | 20 | 21 |
| 3 | 7 | 13 | 10 | 12 |
| 4 | 4 | 12 | 6 | 9 |
| 5 | 3 | 12 | 5 | 8 |
| 6 | 2 | 11 | 5 | 8 |
| 7 | 2 | 11 | 5 | 7 |
| 8 | 2 | 11 | 4 | 7 |

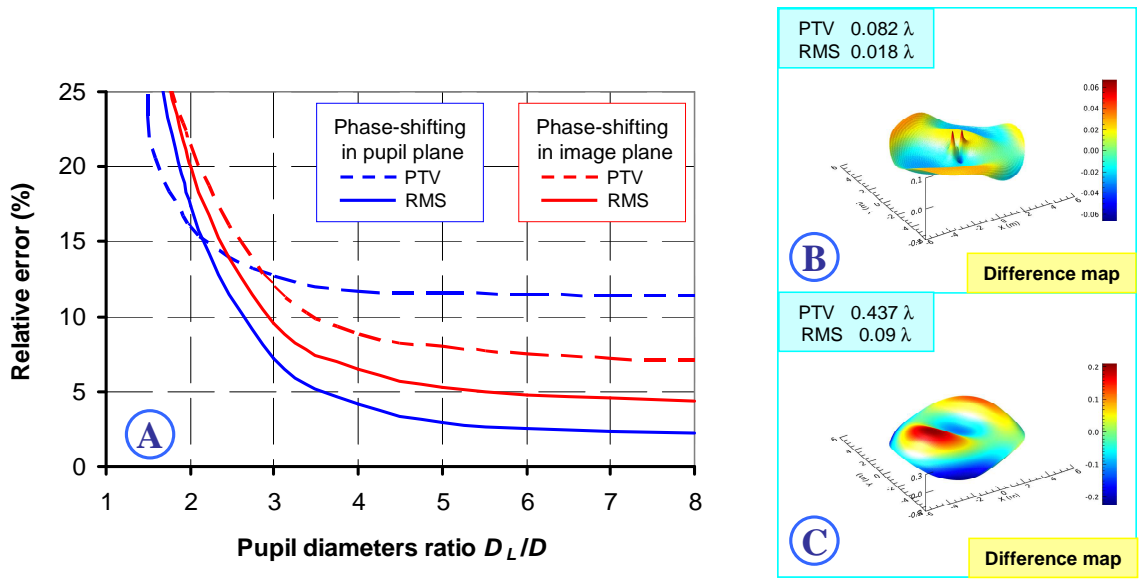

Figure 8: Effect of Lyot stop size on wavefront sensing accuracy. A) Sensitivity curves. B) Error map for phase-shifting in the Lyot stop plane, case $D_L = 2D$. 3) Same phase-shifting process, case of "negative" Lyot stop.

# 5  OPTICAL INPLEMENTATION

Practical implementations of the Phase-shifting telescope interferometer (PSTI) and Phase-shifting Zernike wavefront sensor (PSZ-WFS) have already been proposed and are briefly listed below. We also describe an alternative, dual-mode optical configuration, compatible with coronagraph observation and allowing easy swap between pupil and image plane phase-shifting operation.

**Phase-shifting telescope interferometer (pupil plane)**: The original PSTI optical configuration was presented as an auxiliary telescope followed by relay optics injecting light through a central hole in the secondary mirror of the main telescope [12]. A variant of this design is sketched in Figure 1-3. Ref. [10] also describes alternative configurations where phase-shifting optics are placed after the image plane. Two variants were proposed, either in sequential or single-shot acquisition modes.

**Phase-shifting Zernike wavefront sensor (image plane)**: In the original publications [17-18], phase-shifts are generated in the image plane by moving a single mode optical fiber along the optical axis. The fiber head has been polished and silver coated, thus acting as a micro-mirror. The phase-shifting fiber assembly is set at the focus of a parabola as in a cat's eye delay line arrangement (see Figure 1-4).

**Dual-mode phase-shifting configuration**: In complement to the previous designs, Figure 9 presents a possible implementation of both phase-shifting schemes behind the phase mask of a coronagraph. Firstly, the science and wavefront sensing beams are separated by means of a splitting or dichroics plate located between the collimating optics L2 and the Lyot stop. Only the science beam comprises a Lyot stop, since the Delta approximation in the image plane precludes its utilization for wavefront sensing (see sections 3.2.B and 4.3). In pupil plane phase-shifting mode (Figure 9-A), a small pick-off mirror reflects the central part of the parallel beam at 90 deg, to serve as the reference beam. Two sets of rooftop mirrors M1-M2 and M'1-M'2 are redirecting the beams to a combining mirror M3 that is pierced with a central hole of the same size than the pick-off mirror. The combined beams are then focused onto the WFS detector optically conjugated with the phase mask. The phase-shifts $\phi_m$ are introduced by axial displacement of the rooftop mirror assembly M'1-M'2.

Switching to image plane phase-shifting mode is achieved by inserting three lenses L4, L5 and L'5 into the optical layout, and exchanging the mirror M3 with a beamsplitter BS2 (see Figure 9-A). The lens L4 is dimensioned so that the central lobe of its unaberrated PSF largely covers the full area of the pick-off mirror, which is the basic necessary condition for operating in this mode. One or two detectors are set into a pupil plane optically conjugated with the Lyot stop. Phase-shifts are introduced in the same way as in the pupil plane phase-shifting mode.

**Other solutions**. Less conventional optical implementations could be also envisaged in order to separate the wavefront sensing and coronagraphic arms behind the phase-shifting device, thus decreasing NCPA even more. One may for example consider using a "dichroic" Lyot stop, or an integral field spectrograph having pupil stops of variable sizes.

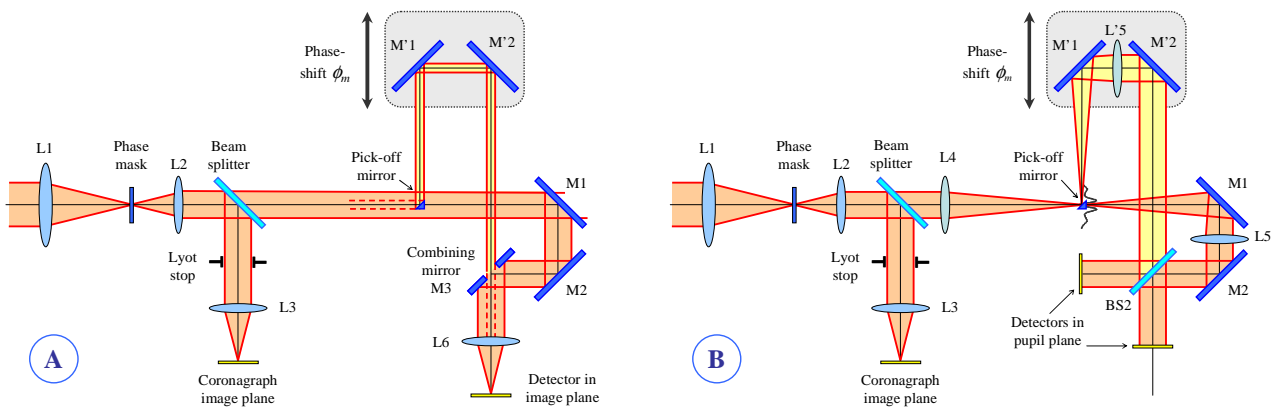

Figure 9: Dual- mode optical configuration for phase-shifting in the pupil plane (A) or in the image plane (B).

## 6    CONCLUSION

In this paper were firstly reviewed the principles of phase-shifting telescope-interferometers or Zernike wavefront sensors envisioned for measuring the wavefront errors of astronomical telescopes, either from pupil or image planes. A Fourier optics theory of those phase-shifting devices was established in view of their application to a coronagraph instrument, from the corrugated entrance wavefront to the measured intensity distributions. Wavefront reconstructions procedures were defined for three typical cases, where phase-shifts are introduced into the telescope pupil plane, the Lyot stop plane, or the coronagraph image plane. They involve two "Dirac" approximations whose validity was discussed. Based on these theoretical models, end-to-end numerical simulations have been carried out in order to evaluate the achievable measurement accuracy in a number of cases, including low and mid-spatial frequency entrance wavefronts and three different types of coronagraphs. The obtained numerical results finally allow answering to some questions raised in section 2.5, especially:

- o   Both phase-shifting concepts (either in pupil or image plane) are well suited to wavefront sensing from behind the phase mask of a coronagraph.

- o   Phase-shifting techniques are not limited to weak aberration measurements but only by $2\pi$-ambiguity inherent to interferometric sensors. Furthermore, they are not restricted to low-order sensing as commonly believed, but also have the capacity to capture mid-spatial frequency WFE residuals.

- o   The most critical parameters are linked to the so-called "Dirac" approximations of sections 3.2 and 3.4. For optimal measurement accuracy, they can be summarized as follows:

    -   Phase-shifting in a pupil plane requires the diameter of the phase-shifting area to be lower or equal than one twentieth of the telescope pupil,

    -   Phase-shifting in an image plane requires it to be lower or equal than one tenth of the Airy disk,

    -   In both cases the aperture of the wavefront sensing optics shall be higher than three times the Lyot stop aperture at least.

In conclusion, phase-shifting techniques could enable accurate wavefront measurements from inside a coronagraph instrument, from its phase mask to the Lyot stop. It could then allow to compensate for most of the non common path aberrations from which these instruments usually suffer. This study also calls for future work about their potential implementation on other types of coronagraphs, such as the shaped pupil, apodized Lyot, or Phase-induced amplitude apodization (PIAA) concepts.

# APPENDIX A. NOMENCLATURE

In this Appendix are compiled most of the shortcut mathematical notations for real and complex bi-dimensional functions employed in sections 3.2 to 3.4, in both pupil and image planes.

**Sections 3.2 and 3.3** (phase-shifts in pupil plane)

- $B_D(P)$    "Pillbox" function of diameter $D$, equal to unity when $\sqrt{x^2 + y^2} \leq R = D/2$ (inside the main pupil), and to zero elsewhere
- $B_d(P)$    Pillbox function of diameter $d = \rho D$, only covering the reference sub-pupil of diameter $d$
- $B_d^D(P)$    Effective area of the main pupil (ring shaped), equal to $B_D(P) - B_d(P)$
- $A_m(P)$    Total complex amplitude into the OXY pupil plane, for the $m^{th}$ phase-shift
- $\hat{B}_D(M')$    Amplitude diffracted by a non aberrated pupil of diameter $D$ into the O'X'Y' image plane. It is the Fourier transform of $B_D(P)$ equal to $2J_1(z)/z$ with $z = \pi D \sqrt{x'^2 + y'^2}/\lambda F$, and $J_1$ is the type-J Bessel function at the first order
- $\hat{B}_d(M')$    Amplitude diffracted by the reference sub-pupil in the X'Y' image plane. Fourier transform of $B_d(P)$ and equal to $2J_1(\rho z)/\rho z$
- $\hat{B}_d^D(M')$    Amplitude diffracted by the non aberrated ring shaped pupil, equal to $\hat{B}_D(M') - \hat{B}_d(M')$
- $\hat{A}'_m(M')$    Total complex amplitude into the O'X'Y' image plane, for the $m^{th}$ phase-shift
- $A''_m(P'')$    Total complex amplitude diffracted into the Lyot plane O"X"Y", for the $m^{th}$ phase-shift
- $B_{D_L}(P'')$    Pillbox function of diameter $D_L$, defining the Lyot stop in plane O"X"Y"
- $\hat{B}_{D_L}(M''')$    Fourier transform of $B_{D_L}(P'')$
- $A'''_m(M''')$    Total complex amplitude into the coronagraphic plane O'''X'''Y''', for the $m^{th}$ phase-shift
- $I'''_m(M''')$    Measured intensity distribution into the coronagraph image plane, for the $m^{th}$ phase-shift
- $\hat{P}(M'')$    Reconstructed complex amplitude into the coronagraphic plane
- $\hat{I}''_m(P'')$    Inverse Fourier transform of $I'''_m(M''')$ retro-propagated in the Lyot stop plane, for the $m^{th}$ phase-shift
- $P(P'')$    Reconstructed complex amplitude into the telescope pupil or Lyot stop planes

**Section 3.4** (additional notations for phase-shifting in image plane)

- $B_{d'}(M')$    Pillbox function of diameter $d'$, equal to unity when $\sqrt{x'^2 + y'^2} = d'/2$ ), and to zero elsewhere. It defines the phase-shifting area In the O'X'Y' plane
- $\hat{B}_{d'}(P'')$    Amplitude diffracted by the phase-shifting area into the Lyot stop plane O"X"Y", Fourier transform of $B_{d'}(M')$
- $I''_m(P'')$    Measured intensity distribution into the pupil or Lyot stop plane, for the $m^{th}$ phase-shift